\documentclass[preprint,review,12pt]{elsarticle}

\usepackage{amssymb}

\journal{Icarus}

\begin{document}

\begin{frontmatter}



\title{Sodium Atoms in the Lunar Exotail: Observed Velocity and Spatial Distributions}


\author{Michael R. Line}
\address{Division of Geological and Planetary Sciences, California Institute of Technology, Pasadena, CA 91106}
\author{E.J. Mierkiewicz}
\address{Department of Physics, University of Wisconsin, Madison, WI, USA.}
\author{R.J. Oliversen}
\address{NASA Goddard Space Flight Center,Greenbelt, MD, USA}
\author{J.K. Wilson}
\address{Space Science Center, University of New Hampshire, Durham, NH, USA.}
\author{L.M. Haffner}
\address{Department of Astronomy, University of Wisconsin, Madison, WI, USA.}
\author{F.L. Roesler}
\address{Department of Physics, University of Wisconsin, Madison, WI, USA.}

\begin{abstract}
The lunar sodium tail extends long distances due to radiation pressure on sodium atoms in the lunar exosphere. Our earlier observations measured the average radial velocity of sodium atoms moving down the lunar tail beyond Earth (i.e., near the anti-lunar point) to be $\sim 12.5$ km/s. Here we use the Wisconsin H-alpha Mapper to obtain the first kinematically resolved maps of the intensity and velocity distribution of this emission over a $15 \times 15 \deg$ region on the sky near the anti-lunar point. We present both spatially and spectrally resolved observations obtained over four nights bracketing new Moon in October 2007. The spatial distribution of the sodium atoms is elongated along the ecliptic with the location of the peak intensity drifting $3 \deg$ east along the ecliptic per night. Preliminary modeling results suggest the spatial and velocity distributions in the sodium exotail are sensitive to the near surface lunar sodium velocity distribution. Future observations of this sort along with detailed modeling offer new opportunities to describe the time history of lunar surface sputtering over several days.
\end{abstract}

\end{frontmatter}


\section{Introduction}

The Moon is known to have a trace atmosphere of helium (He), argon (Ar), sodium (Na), potassium (K) and other trace species [see e.g., {\em Stern} 1999]; however its tenuous nature makes remote observations of elements other than the alkalis difficult. Sodium ``D-line'' emission at 5895.924 \AA\ (D1) and 5889.950 \AA\ (D2) has been used since the late 1980s to observe and interpret the morphology of the lunar sodium atmosphere beginning with its detection by {\em Potter and Morgan} [1988] and {\em Tyler et al.} [1988]. Likely source mechanisms are: thermal desorption, photo-desorption, ion sputtering and meteoric impact ablation. The relative importance of these mechanisms remains uncertain, both with regard to spatial and to temporal trends. Once released, sputtered gases in the lunar atmosphere can be pulled back to the regolith by gravity, escape to space, get pushed away by solar radiation pressure, or become photoionized and swept away by the solar wind.

{\em Mendillo et al.} [1991] obtained the first broadband imaging observations (D1 + D2) of the extended lunar sodium atmosphere, observing emission out to $\sim$5 lunar radii ($R_{m}$) on the dayside, and out to 15--20 $R_{m}$ in a ``tail-like'' structure on the nightside. The lunar sodium tail is now known to extend to great distances (many hundreds, and perhaps thousands, of lunar radii) due to the strong influence of the Sun's radiation pressure on Na atoms in the lunar exosphere [{\em Wilson et al.}, 1999]. Near new Moon phase, the extended lunar sodium tail can be observed as it sweeps over the Earth and is gravitationally-focused into a visible sodium ``spot'' in the anti-solar direction [{\em Smith et al.}, 1999; {\em Matta et al.}, 2009]. Refer to Figure 1. 

Velocity resolved observations of the extended lunar sodium tail were first obtained by {\em Mierkiewicz et al.} [2006a] using a large aperture (15 cm) double etalon Fabry-Perot spectrometer with $\sim 3.5$ km/s spectral resolution at 5890 \AA; see {\em Mierkiewicz et al.} [2006b] for further instrument details. Observations were made within 2--14 hours of new Moon from the Pine Bluff Observatory (PBO), Wisconsin, on 29 March 2006, 27 April 2006 and 28 April 2006. The average observed radial velocity of the lunar sodium tail in the vicinity of the anti-solar/lunar point for the three nights was $\sim 12.5$ km/s (from geocentric zero). This velocity is consistent with sodium atoms escaping from the Moon and being accelerated by radiation pressure for 2+ days. In some cases the line profile appeared asymmetric, with lunar sodium emission well in excess of 12.5 km/s. 

In this paper we report new results using the unique mapping capabilities of the Wisconsin H-alpha Mapper (WHAM), where we have traded spectral resolution in favor of increased sensitivity and spatial resolution. 

\section{Instrumentation}

WHAM was built to map the distribution and kinematics of ionized gas in our Galaxy [{\em Haffner et al.}, 2003]. Here we leverage WHAM's unique combination of high sensitivity, spectral resolution and automated pointing capabilities to map Na emission in the extended lunar sodium tail. At the time of these observations WHAM was located at Kitt Peak Observatory, AZ. 

Similar in design to the PBO Fabry-Perot used in our earlier work, WHAM is a large aperture (15 cm) double etalon Fabry-Perot coupled to a siderostat with a circular $1 \deg$ field-of-view (FOV) (compared with $1.5 \deg$ for PBO) . WHAM has a resolving power of 25,000, covering a 200 km/s spectral region with 12 km/s spectral resolution at  5890 \AA\ [{\em Haffner et al.}, 2003]. 

\section{Observations \& Reduction}

Using WHAM we have mapped the intensity and velocity distribution of the extended lunar sodium tail over a $15 \times 15 \deg$ region near the anti-lunar point with $1 \deg$ spatial resolution. Observations were made during 4 nights bracketing the 11 October 2007 (5:01 UT) new moon period. The automated pointing capability of WHAM was used to build a map of the lunar Na emission by rastering WHAM's $1 \deg$ circular FOV in a ``block'' surrounding the anti-lunar point; see Figure 2. The number of $1 \deg$ pointings per block was between 121 and 256. The exposure time for each observation was 120 s; a map was generated in approximately 6 hours. These observations provide the first kinematically resolved maps of the extended lunar sodium tail observed in the anti-lunar direction. 

Individual WHAM spectra were reduced with a four-component model: one Gaussian component for an atmospheric OH feature near -90 km/s from geocentric zero, a second Gaussian component for an unidentified feature near -70 km/s, a Voigt profile for the terrestrial sodium emission at 0 km/s and a Gaussian component for the lunar sodium emission near 13 km/s; refer to Figure 2a. 

Due to the partial blending of the terrestrial and lunar Na emission, the component fitting of the WHAM data required two steps. First, a constrained fit was applied to the data in which the Doppler shift of the lunar emission with respect to the terrestrial sodium sky glow line was fixed at 12.5 km/s based on our experience with the higher resolution PBO observations. Next, after a best-fit solution was obtained, the Doppler separation of the lunar emission was freed and the fitting routine was run again. In all cases, fit components were convolved with an instrumental profile and iterated, subject to the above constraints, to produce a least squares, best-fit to the data using the VoigtFit code of {\em Woodward} [private communication, 2012]. 

Briefly, VoigtFit is a parameter estimation package for the analysis of spectral data. VoigtFit uses the Levenberg-Marquardt method of estimating parameters by minimizing chi-square using a hybrid of the steepest-decent and quadratic (Hessian) methods; the Levenberg-Marquardt method is described in {\em Numerical Recipes} [{\em Press et al.}, 1986]. Of particular importance here is VoigtFit's ability to: 1) analytically link parameters, which allows overlapping or faint lines to be fit without unnecessary free parameters, and 2) incorporate an empirical instrumental profile into the fitting process (obtained from a spectrum of a thorium emission line from a hollow cathode lamp) [{\em Woodward}, private communication, 2012].

After fitting, we plot the lunar sodium emission spectra arranged according to their positions on the sky (Figure \ref{fig:data_reduce}b).   The intensity (i.e., the integrated area under the emission line, converted to Rayleighs, where 1R $= 10^{6}/4\pi$ photons cm$^{-2}$ s$^{-1}$ str$^{-1}$) of the lunar sodium emission is then converted into colored beams representing WHAM's $1 \deg$ FOV (Figure \ref{fig:data_reduce}c).  These beams are then smoothed (Figure \ref{fig:data_reduce}d).   We also generate a spatial map for the Doppler widths (full-width-at-half-maximum) of the lunar sodium emission lines.  

Intensity calibration is based on the surface brightness of NGC 7000, the ``North American Nebula'' (coordinates $\alpha_{2000} = 20.97$ h, $\delta_{2000} = 44.59 \deg$). The NGC 7000 hydrogen Balmer $\alpha$ (6563 \AA) surface brightness is $\sim 800$ R over WHAM's $1 \deg$ FOV [{\em Haffner et al.}, 2003]. Sodium D2 line intensities are based on the assumption that WHAM's efficiency is unchanged between 6563 \AA\ and 5890 \AA, and a measured filter transmission ratio of T(5890 \AA)/T(6563 \AA) $\sim 1$. We estimate a 25\% uncertainty in our sodium D2 absolute intensity calibration. 

\section{Results}

Spectra for the brightest beam for each night are given in Figure 3 and Table 1. Intensity and Doppler width maps for all pointings are given in Figure 4. 

In order to determine the sky background, we mapped a region of the sky $60 \deg$ east of the anti-lunar point at $-20.15 \deg$ ecliptic latitude and $57.82 \deg$ ecliptic longitude. This off-direction dataset was processed with the Gaussian fitting procedure used in the analysis of the on-direction datasets as described in Section 3. We found intensities greater than 0.7 R to clearly be lunar Na emission (see Figure 4).

In what follows we present a basic description of the observations for each night.\\

\noindent
{\bf 10 October 2007:} Observations were taken 22 to 16 hours before new Moon (first column of Figure 4), 169 one $\deg$ pointings. The intensity distribution is elongated along the ecliptic with the location of the peak intensity to the southwest of the antisolar point.  The Doppler width distribution appears to peak to the northeast of the brightest emission, and is also elongated along the ecliptic.  The broadest emission is also the faintest emission for this night. The brightest emission occurs approximately $10.8 \deg$ from the antisolar point, southwest along the ecliptic.\\

\noindent
{\bf 11 October 2007:} Observations were taken 1 hour before to 7 hours after new Moon (second column of Figure 4), 225 one $\deg$ pointings. The intensity distribution is nearly axially symmetric; the Sun, Moon and Earth are nearly aligned, and we are looking almost directly down the lunar tail.  As with the night of 10 October, the peak intensity is to the southwest of the antisolar point. The broadest emission for this night is near the brightest emission. The peak emission occurs $5.8 \deg$ southwest of the anti-solar point.\\

\noindent
{\bf 12 October 2007:} Observations were taken 23 to 32 hours after new Moon (third column of Figure 4), 256 one $\deg$ pointings. The intensity distribution again appears to be elongated along the ecliptic with the brightest emission occurring closer to the anti-solar point and the fainter emission occurring farther to the southwest.  From the Doppler width map, there is a large region where the widths are roughly 15-20 km/s, with the broadest emission falling roughly in the same location as the brightest emission, $2.7 \deg$ from the anti-solar point. \\

\noindent
{\bf 13 October 2007:} Observations were taken 48 to 52 hours after new Moon (fourth column of Figure 4), 121 one $\deg$ pointings. The emission is extremely faint and remains elongated along the ecliptic.  The broadest emission, upwards of 30 km/s, for this night appears to peak to the southwest of the brightest emission.  The brightest emission is  $1.8 \deg$ to the southwest of the anti-solar point.
   
\section{Sample Model Runs}

Here we present a sample set of data/model comparisons using the numerical Monte-Carlo lunar exospheric simulation model of {\em Wilson et al.} [1999; 2003]. We include these sample model runs to illustrate the potential of future model/data analysis.  

The {\em Wilson et al.} [1999; 2003] Monte-Carlo model uses fourth-order Runga-Kutta integration to compute the accelerations and positions of $\sim10^{6}$ lunar exospheric sodium atoms due to gravitational effects and solar radiation pressure (a function of the atom's heliocentric distance and velocity); radiation pressure shadowing by the Moon and Earth is included in the model. The model accounts for the motion of the Moon around the Earth and the Earth around the Sun.   Atom loss due to impact onto the Earth and by photoionization is also taken into account. A photoionization lifetime of 47 hours [{\em Heubner et al.}, 1992] was used in all model runs presented here. 

In each model run, sodium atoms were randomly ejected from the Moon beginning 4 days before the desired simulation snapshot to ensure that the lunar exosphere/tail was well populated.   The initial velocities at the Moon in the x, y and z directions were randomly determined from 0.1 km/s velocity bins between 2.1 and 2.6 km/s.  The atom's ejection angle was set by the x, y and z components, resulting in an isotropic angular distribution. The relative fraction of particles in each velocity bin was a free model parameter (refer to Section 5.2). Because each of our maps was built over a $\sim 6$ hour observing interval, a number of simulation snapshots (one per hour of observation) were computed for each observing interval and coadded to simulate the smearing inherent in our observations as the tail moves throughout the night.  The sodium ejection rate was set to $1\times10^{22}$ atoms/s [{\em Wilson et al.}, 1999]. Note, although the ejection rate primarily controls the intensity of the sodium tail and therefore may be treated as a retrieved model parameter, at this time we are not attempting to derive the ejection rate from our data. Here we remain focused on the initial velocity distributions and their effect on the morphology of the extended lunar tail.

In the sections that follow, we compare the model and data both spatially (Section 5.1) and spectrally (Section 5.2).

\subsection{Spatial Map Comparisons}
Figure 5 shows the tail geometry at the time of our observations for each night, along with the model intensity distribution projected onto the sky, and our data for comparison.  Figure 6 shows the tail line-of-sight velocities of the atoms with respect to Earth. These model runs are for a flat initial velocity distribution (see Figure 7 \& Section 5.2). The line-of-sight through the tail for the anti-solar direction (dotted line Figures 5 \& 6) and the direction of the brightest emission (solid line in Figures 5 \& 6) are shown to help visualize how the tail appears in projection onto the plane of the nightsky.

The dense core of the tail (see Figure 5) is due to gravitational focusing of the sodium atoms by the Earth, and hence the brightest emission tends to occur while looking down this portion of the tail. As seen in the data, and with agreement in the model runs, the brightest emission always occurs to the west of the anti-solar point and moves eastward over the course of our four nights of observations. This eastward drift is also seen as a decrease in the angle between the anti-solar line-of-sight and the brightest emission line-of-sight over our four nights of observations (see Figure 5). When the Moon is to the east of the Earth-Sun line, as it is on the night of 10 October, the sodium atoms are deflected to the west.  This deflection is responsible for the large westward location of the brightest emission observed on this night. As the Moon moves to the west of the Earth-Sun line (see Figure 5), the Earth deflects the lunar sodium atoms to the east.  This explains why the brightest emission on the sky appears closer to the anti-solar point on the nights of 12 and 13 October.

In both the data and the model images in Figure 5 a dark spot appears near the anti-solar point. It's unclear to what extent the Earth's gravitational deflection may contribute to this dark spot, but the Earth's shadow must play a role as it reduces/prevents illumination by the Sun.

\subsection{Spectral Comparisons}
Spectral averaging is a useful metric to explore how the velocity distribution evolves from night-to-night.  Both the data and model averages were normalized for these comparisons. Model calculations were binned to a $1 \deg$ spatial resolution to match the $1 \deg$ spatial resolution of the WHAM spectrometer. Model spectra are generated with significantly higher velocity resolution than WHAM's spectral resolution of 12 km/s. As such, model output was convolved with a WHAM instrumental function in order to facilitate direct data/model comparisons.   
 
We investigated how variations in the initial sodium particle velocity distribution at the Moon manifest themselves in our observed line profiles from the 11 October 2007 dataset. The solid line in Figure 7 is an average line profile for all WHAM lunar sodium observations obtained on 11 October 2007 after removal of the fitted terrestrial sodium emission from the original data. The line profile was normalized to the peak intensity, ignoring the residual noise spikes associated with sodium terrestrial line subtraction. The dashed line in Figure 7 is the corresponding model run average, normalized to its peak intensity. The model run average spectra were computed from single model simulation snapshots for the time of new Moon.

Three different initial velocity distributions were used in our data/model comparison: slow (initial velocities falling between 2.1 and 2.15 km/s), fast (initial velocities falling between 2.56 and 2.6 km/s) and flat (an equal number of particles leaving the Moon with velocities of 2.1--2.6 km/s).   These velocity distributions are based on the work of {\em Wilson et al.} [1999; 2003]. Although the initial velocity distribution was not fully constrained by {\em Wilson et al.} [1999; 2003], the 2003 paper did show that there is a significant contribution to the extended lunar Na tail from source speeds in the range of 2.1--2.4 km/s. Even though more sophisticated velocity distributions, such as a Maxwell-Boltzman distribution, could be used, here we simply wanted to explore population extremes in order to determine the feasibility of this and future data to constrain the lunar surface velocity distribution. 

Inspection of Figure 7 indicates that the shape of the observed and modeled line profiles in the anti-lunar direction near the time of new Moon are sensitive to the initial velocity distribution of the sodium atoms leaving the Moon. A fast initial velocity distribution results in a broader average spectrum whereas a slow initial distribution produces a much narrower spectrum. It is important to emphasize that changes in the lunar source rates $\sim 2$ days prior to the observations (when the observed Na atoms were being released from the lunar surface) could produce similar signatures, as would changes in the sodium photoionization lifetime due to solar activity. The flat velocity distribution produces a spectrum closest to that observed; see Figure 7, lower right. 

Figure 8 shows model runs for the other three nights of observations (together with 11 October), using the flat initial velocity distribution. As with Figure 7, the solid line in each panel of Figure 8 is an average line profile for all WHAM lunar sodium observations obtained on that night after removal of the fitted terrestrial sodium emission from the original data. The line profiles were normalized to the peak intensity, ignoring the residual noise spikes associated with sodium terrestrial line subtraction; the dashed line is the corresponding model run average, normalized to its peak intensity. In this case model spectra were computed and averaged over the observation time ranges for each night (refer to Figure 4).

In both the data and model, the night of 13 October displays the broadest emission. Geometrically, as seen in Figure 6, our observations on this night have the best view down the tail, unobscured by the dense core, sampling the greatest range of radial velocities. This effect manifests itself in the Doppler width maps of Figure 4 (lower panel). The broadest emission observed on 13 October is $\sim 30$ km/s; sodium atoms achieve these velocities $\sim$1.5 million km (3.75 times the Earth-Moon distance) downwind from the Earth.

\section{Conclusions}

Over the course of 4 nights ($\sim$70 hours) of observations, the peak of the intensity distribution (see Figure 4 and Table 1) drifted east along the ecliptic a total $\sim 11.5 \deg$ (an average of $0.16 \deg$ per hour), consistent with the 3--4 $\deg$ eastward drift per day observed by {\em Smith et al.} [1999]. For reference, the lunar motion is about $0.5 \deg$ per hour eastward.  The observed eastward drift of the brightest emission is due to a combination of the Moon's orbital motion and the gravitational deflection of sodium atoms by the Earth. The brightest emissions occurred on the nights of 11 and 12 October as the observation geometry on these nights presents the most direct look down the gravitationally focused part of the lunar tail (and hence the maximum column emission). 

The broadest line profiles were detected on the nights away from new Moon, occurring northeast of the peak intensity for the night preceding new Moon, and to the southwest following new Moon.  Our preliminary modeling efforts suggest that the changes in the observed morphology are related to viewing geometry as the tail sweeps past the Earth (see Figures 5 \& 6). 

At new Moon the nearby bright ``core'' of Na atoms, recently gravitationally focused by the Earth [{\em Wilson et al.}, 1999], dominates the tailÕs appearance giving it a nearly axisymmetric core emission at 12.5 km/s, obscuring the dimmer signal of the older (and faster) more distant atoms. Before and after new Moon, however, the Earth's gravity more strongly influences atoms at one outer edge of the sodium tail and the tail is observed off-axis, leaving relatively more atoms in an extended diffuse ``un-focused'' tail away from the 12.5 km/s core emission and a correspondingly larger influence on the Doppler width observed in our data. Refer to Figures 4--6. 

Our sample model runs and recent work by {\em Lee et al.} [2011] confirm that velocity resolved observations and spatial mapping of the extended lunar tail offer new opportunities to describe the time history of lunar surface sputtering over several days, and set constraints for models of exospheric source mechanisms and their variabilities. 

\section{Acknowledgments}
The authors thank M. Mendillo and R. Reynolds for their valuable assistance as well as K. Nordsieck for providing the sodium filter. We  thank all the members of the WHAM collaboration, in particular K. Jaehnig, A. Hill, G. Madsen and K. Barger. Finally, we thank the National Solar Observatory mountain support staff, C. Plymate and E. Galayda for their support and hosting us during the WHAM observations. M. Line's involvement as an undergraduate at Wisconsin was partially supported by a UW-Madison Hilldale Undergraduate Fellowship. This work was also funded by NASA award NNX11AE38G. WHAM construction and operations were primarily supported by the National Science Foundation; in particular, the use of WHAM described here was partially supported by awards AST-0607512 and AST-1108911.

\label{}








\newpage 
 
\begin{table}[h]
\caption{October 2007 WHAM extended lunar sodium tail observations.  The right ascension ($\alpha$), declination ($\delta$), Intensity (I), Doppler Shift ($\Delta v_s$) and Doppler Width ($\Delta v_w$, fwhm) are given for the brightest pointing for each night.  New Moon occurred on 11 October at 5:01 UT.}
\begin{tabular}{cccccc}
Obs. Time/Date (UT) &  $\alpha$ (hours) &$\delta$ ($\deg$) &I (R) &$\Delta v_s$ (km/s) &$\Delta v_w$(km/s)  \\
\hline
08:30 10 Oct. 2011& 0.35 & 1.83& 2.2 & 9.37 & 7.19 \\
05:51 11 Oct. 2011& 0.73 &  3.96& 6.3&11.96 & 11.29 \\
08:02 12 Oct. 2011& 0.93 & 5.96& 8.7 &17.97 & 27.78\\
06:48 13 Oct. 2011& 1.05 &6.96 & 3.5 &30.39 & 40.36\\
\hline
\end{tabular}
\end{table}

\newpage
\begin{figure*}[h]
\begin{center}
\includegraphics[height=1.in,width=!, angle=0]{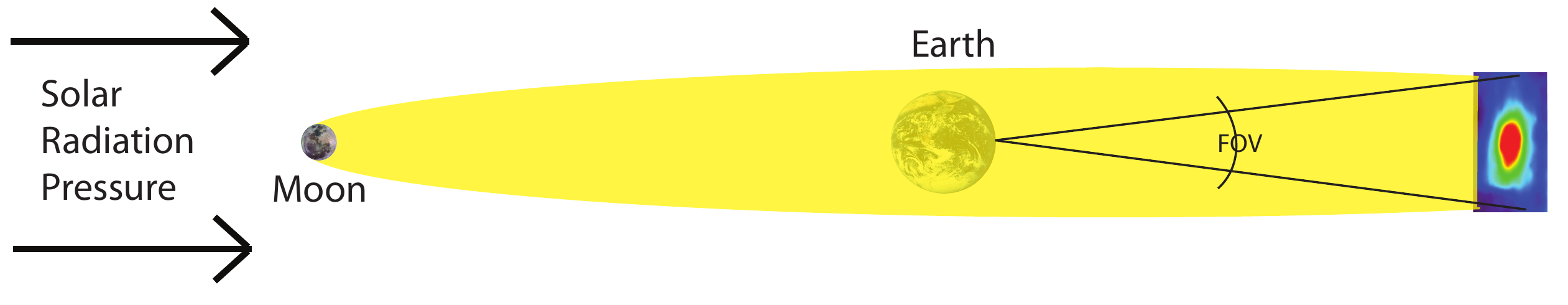}
\end{center}
\caption{ \label{fig:geometry}A cartoon depicting the geometry of the extended lunar sodium tail.  Sodium observations are made in the anti-lunar direction near New Moon, looking down the lunar tail as it moves beyond the Earth, along the Sun-Moon-Earth line (after {\em Wilson et al.} [1999])}
\end{figure*} 
\newpage
\begin{figure*}[h]
\begin{center}
\includegraphics[height=4.5in,width=!, angle=0]{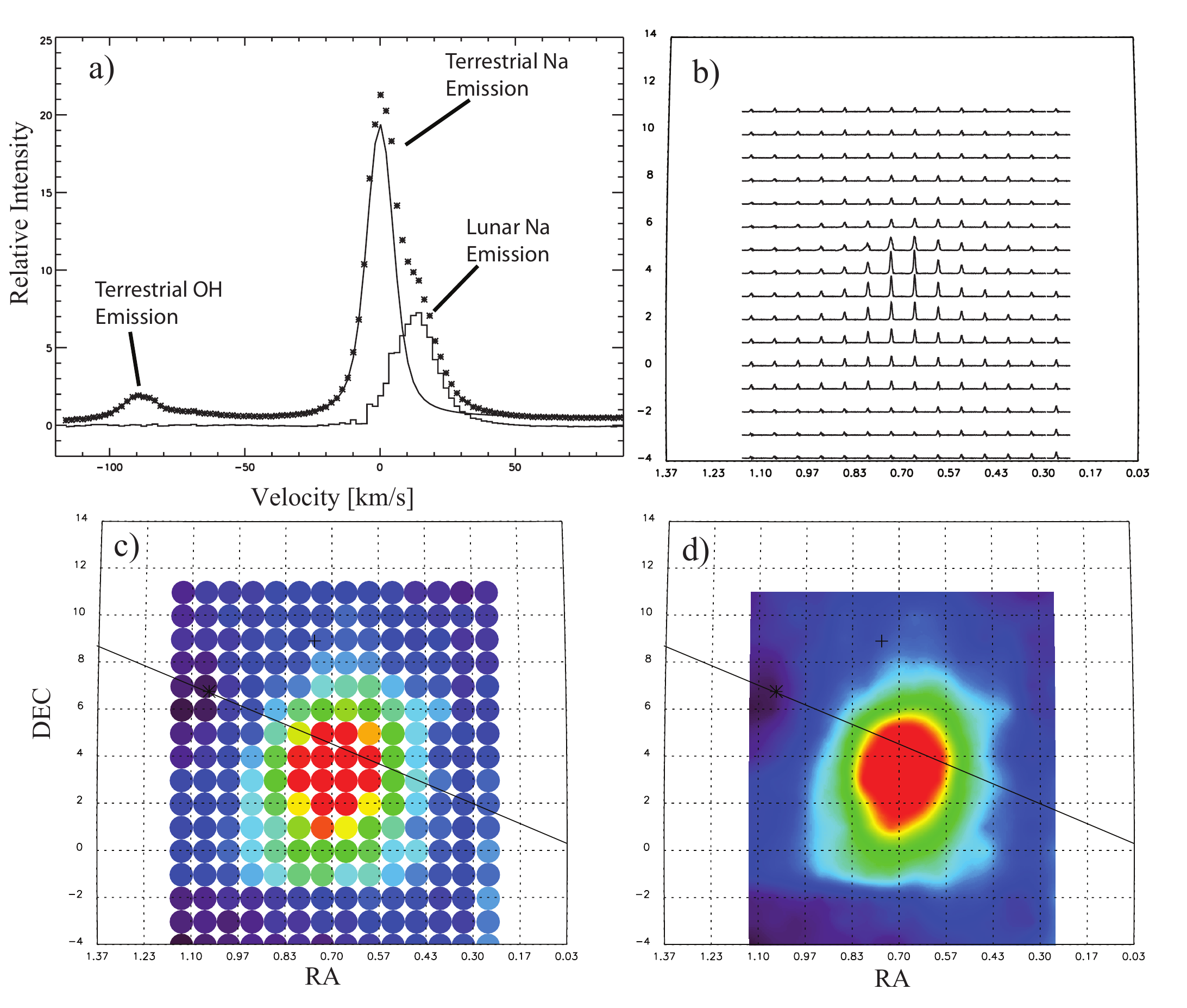}
\end{center}
\caption{ \label{fig:data_reduce} Data reduction procedure. (a) A typical spectrum on 11 October 2007 showing both terrestrial (0 km/s) and Doppler shifted lunar sodium emission near 12.5 km/s. A Gaussian component fit is used to isolate and remove the terrestrial emission and to track the relative intensity and Doppler width of the lunar emission. (b) Spatial variation of the lunar sodium emission on the sky (as a function of right ascension (hours) and declination ($\deg$)).  (c) The intensity of the lunar emission is converted into a colored beam map (each beam is $1 \deg$ on the sky). (d) Smoothed version of (c). A similar procedure is used to visualize the spatial distribution of the Doppler width of the lunar emission. The anti lunar point is indicated by the +, the anti solar point is indicated by the *, and the solid black line is the ecliptic.}
\end{figure*} 
\newpage
\begin{figure*}[h]
\begin{center}
\includegraphics[height=4.in,width=!, angle=0]{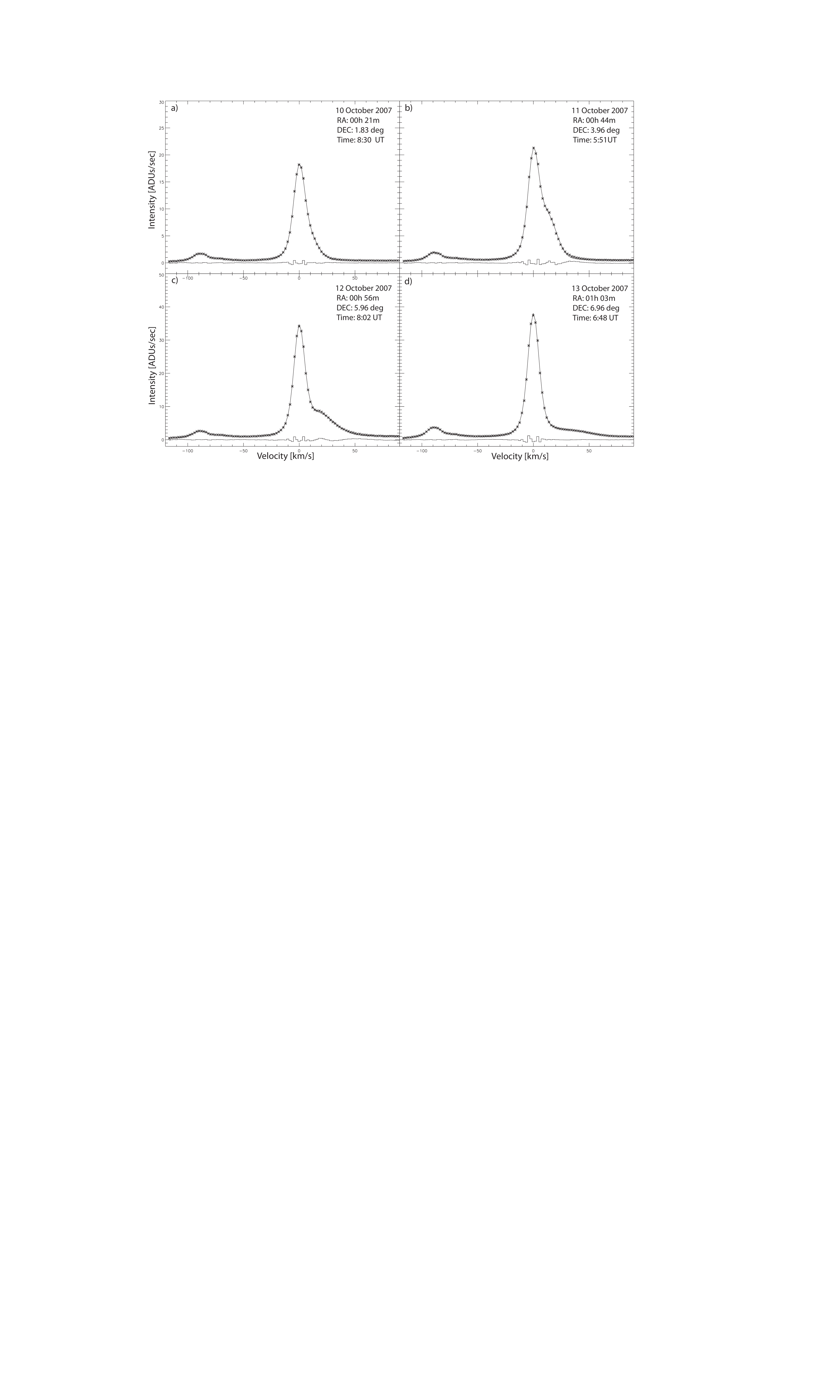}
\end{center}
\caption{\label{fig:brightest} The brightest lunar emission spectra for each of the four October nights.  In all cases the lunar emission is blended with the terrestrial emission; the decomposition of (b) is shown in Figure 2. The lunar emission is redward of the terrestrial emission (near 12.5 km/s).  The dates, times and celestial coordinates are given in each panel.  The points are the data and the solid black curves are the four-component Gaussian fits. The residuals of the fit are also included, centered on zero.}
\end{figure*} 
\begin{figure*}[h]
\begin{center}
\includegraphics[height=3.5 in,width=!, angle=0]{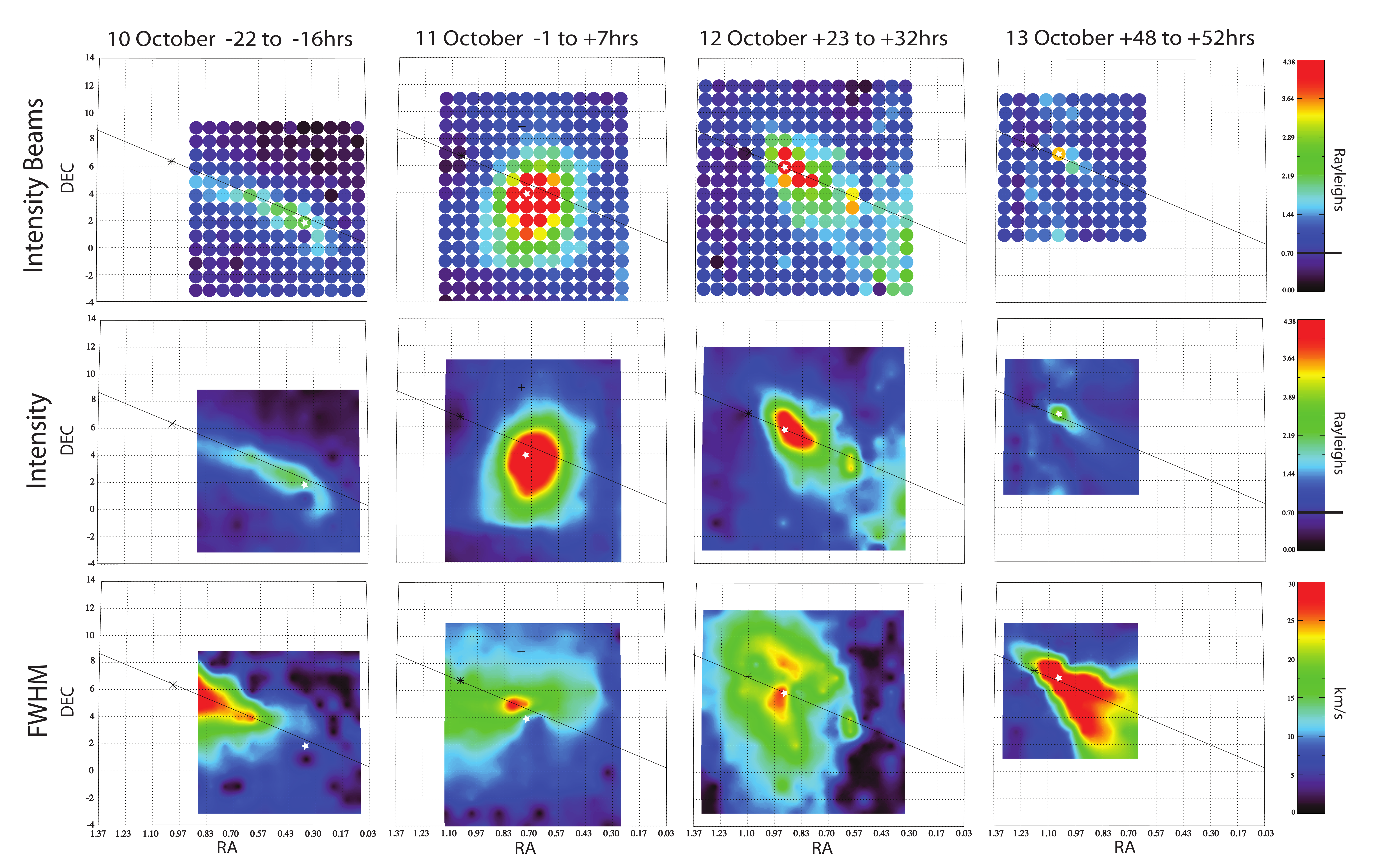}
\end{center}
\caption{ \label{fig:observations}October WHAM Observations.  
(top) 
Regularly gridded beam maps of the lunar sodium emission for each night; right ascension (RA) is given in hours, declination (dec) is in degrees. The anti lunar point is indicated by the + (in the downwind direction), the anti solar point is indicated by the *.  The brightest beam observed for each evening is indicated by a white star.  The black line represents the ecliptic. (middle and bottom) Smoothed map of the spatial variation in the intensity and Doppler width (red corresponds to broader emission) of a single Gaussian fit to the lunar sodium emission. The times relative to new Moon for each night are shown at the top of the Figure.  The solid line on the intensity color bars denotes our confidence limit; colors/intensities above this value are lunar sodium emission (see text).}
\end{figure*}
\newpage   
\begin{figure*}[h]
\begin{center}
\includegraphics[height=6in,width=!]{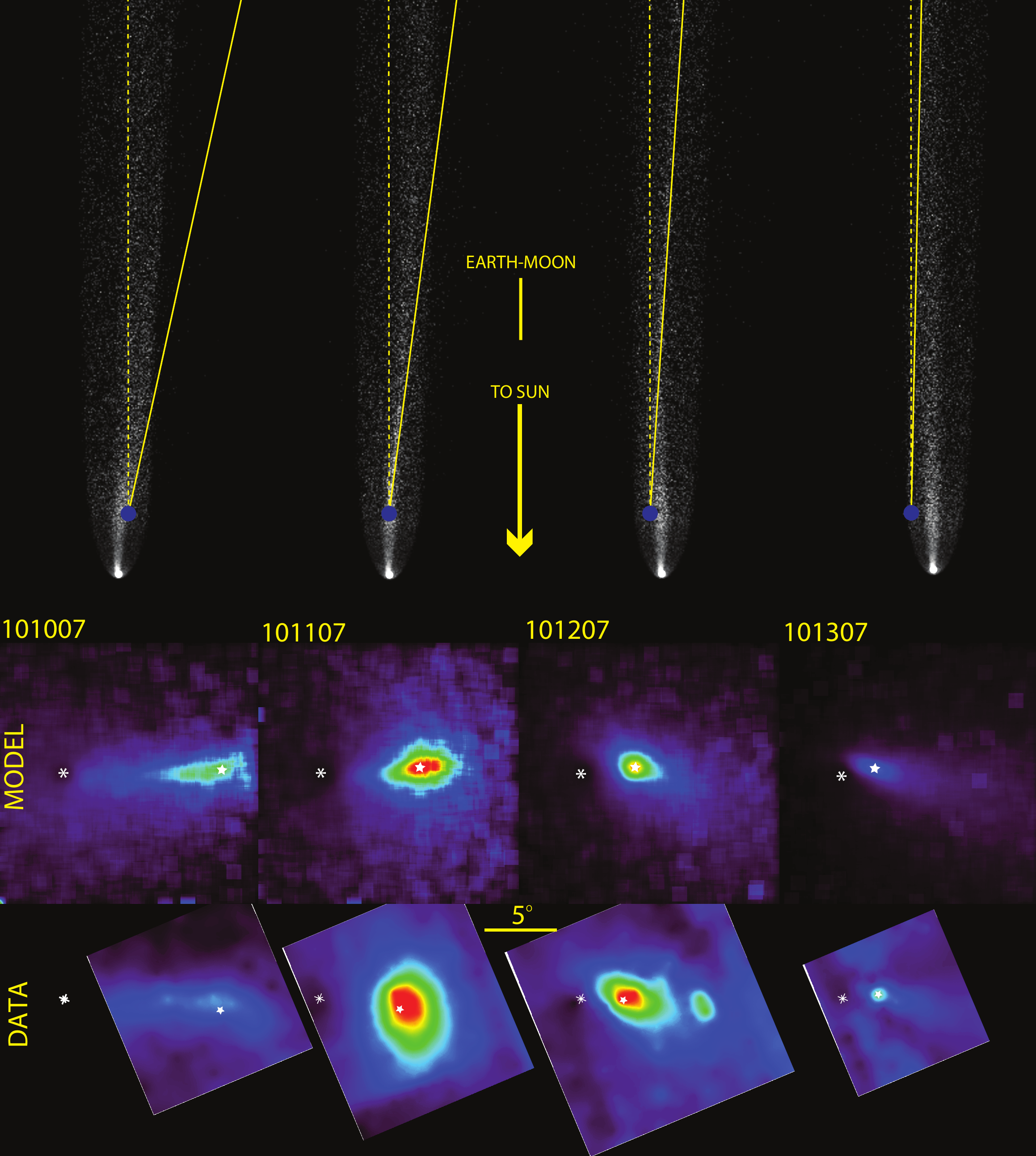}
\end{center}
\caption
{
October 2007 observations (bottom) and data-model comparisons (top panels). (top) Model tail particle density maps looking down on the ecliptic plane from above; the blue dot represents Earth; the white dot represents the Moon (the Earth and Moon are not to scale, but the distance between their centers is to scale). The dashed line is the line-of-sight in the anti-solar direction; the solid line is the line-of-sight toward the brightest pointing (indicated by the star in the middle and bottom panels). (middle) Model and (bottom) data emission maps on the sky in ecliptic coordinates, normalized to the brightest model emission (red indicates bright emission, blue faint). The asterisk locates the anti-solar point. The size of each image is $18 \times 18 \deg$. North is up, west is to the right. 
}
\end{figure*}

\begin{figure*}[h]
\begin{center}
\includegraphics[height=4.75in,width=!]{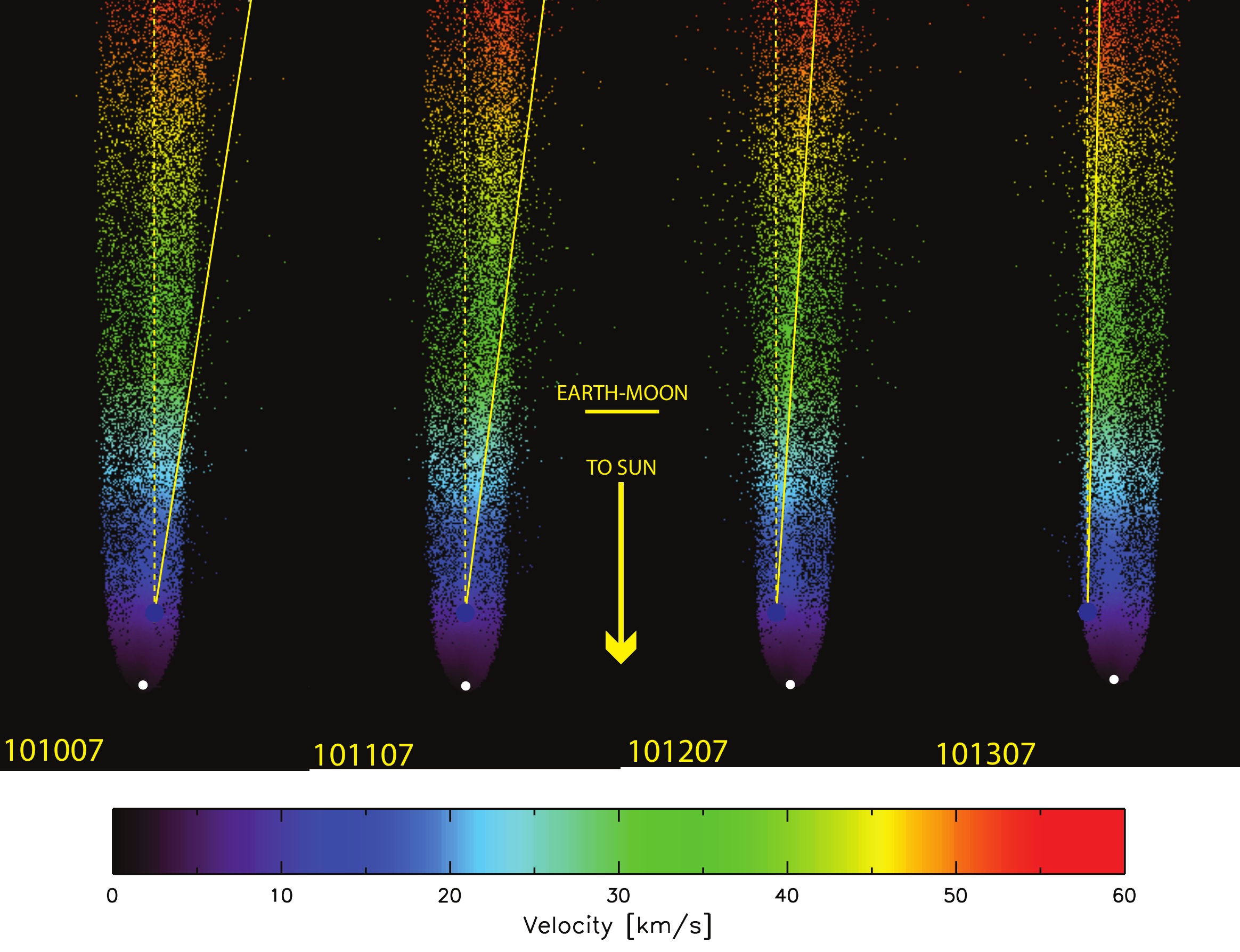}
\end{center}
\caption
{Model tail particle density maps looking down on the ecliptic plane from above, color coded for line-of-sight velocities. As in Figure 5, the lines-of-sight for the anti-solar point (dotted line) and the brightest emission (solid line) are shown.  The small white dot represents the Moon, the larger white dot represents the Earth (not to scale).  Red indicates fast moving particles ($\sim 60$ km/s) and blue slow moving particles.  The Earth-Moon distance is given by the horizontal scale bar labeled Earth-Moon.  Use this Figure along with the Doppler width maps of Figure 4 to aid the interpretation of the observed velocity distribution.}
\end{figure*}

\begin{figure*}[h]
\begin{center}
\includegraphics[height=4.25in,width=!]{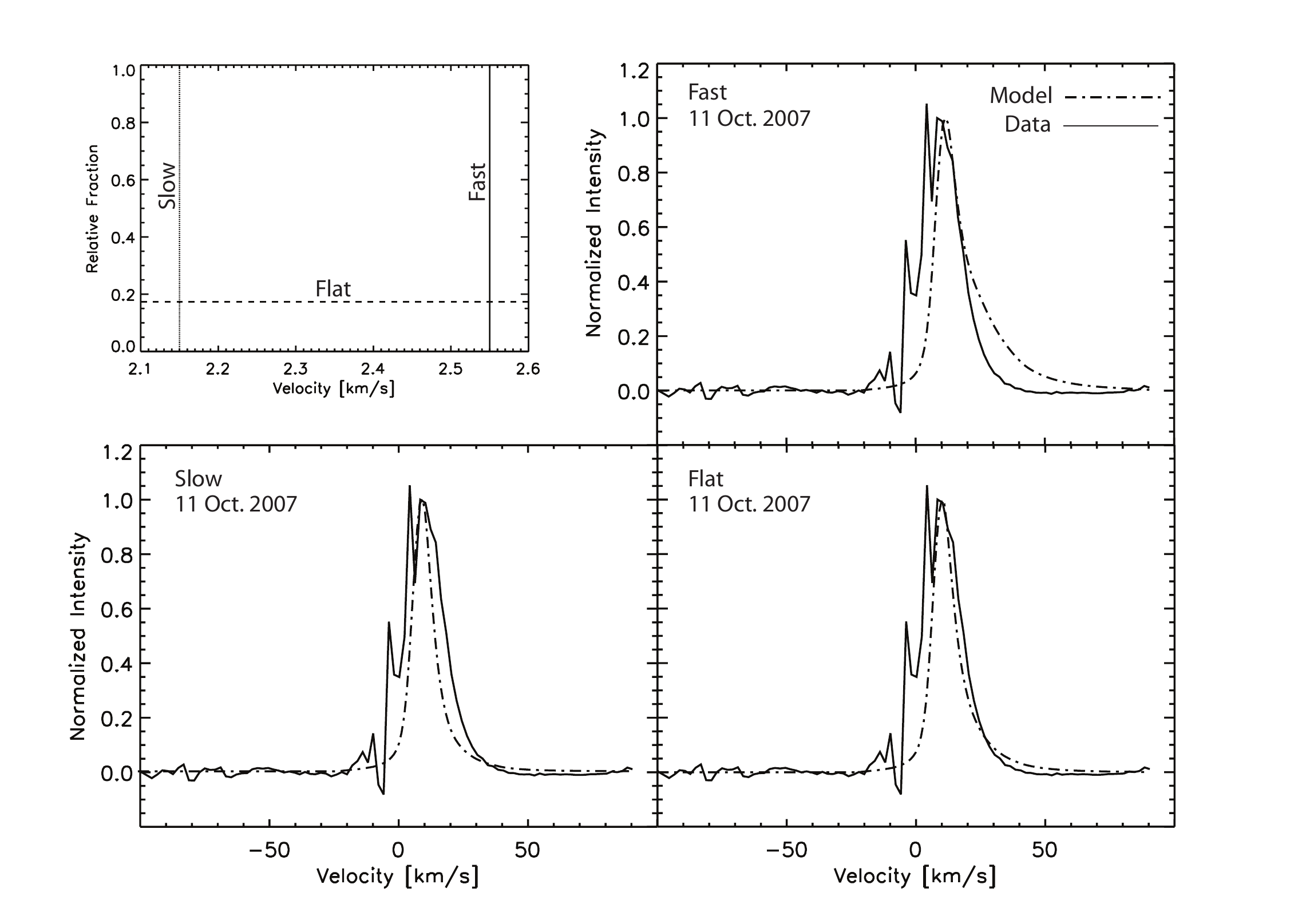}
\end{center}
\caption
{
Data/model line profile comparisons using 3 different initial velocity distributions: slow, flat, and fast. (inset) The relative fraction of the particles that fall within a particular velocity interval. The ``slow'' velocity distribution represents particles leaving the Moon with initial velocities that fall between 2.1--2.15 km/s.  In the ``fast'' case, initial velocities fall between 2.56--2.6 km/s.  The ``flat'' velocity distribution represents an initial distribution with a range of velocities between 2.1--2.6 km/s. (top right \& bottom) Model and data (11 Oct. 2007) comparisons using the fast, slow and flat initial velocity distributions. The noise on the blue wing of the profile is due to residuals in the terrestrial sodium sky glow line subtraction.
}
\end{figure*}

\newpage
\begin{figure*}[h]
\begin{center}
\includegraphics[height=4.in,width=!]{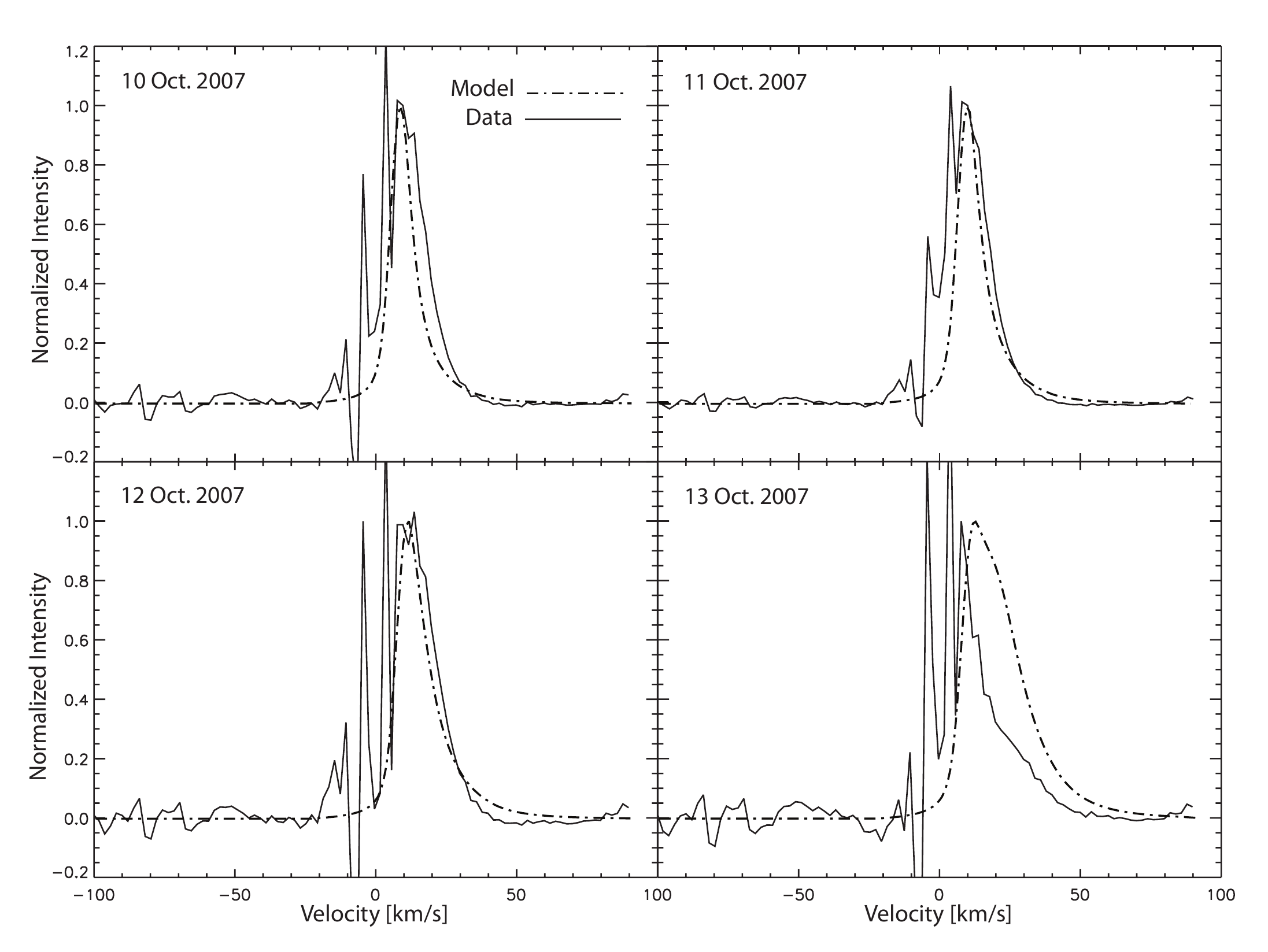}
\end{center}
\caption
{
A comparison of the average spectra for each of the four October nights from WHAM and the average spectra generated for each of the October nights via the ``flat'' model (see Figure 7). The solid line represents the WHAM data and the dashed line represents the spectra produced by the model. In this case model spectra were computed and averaged over the observation time ranges for each night (see Figure 4). 
}
\end{figure*}

\end{document}